\documentstyle[11pt]{article}\textheight 230mm\textwidth 150mm
            \newcommand{\be}{\begin{eqnarray}}
            \newcommand{\ee}{\end{eqnarray}}
            \newcommand{\eel}[1]{\label{#1}\end{eqnarray}}
            \newcommand{\vb}{{\cal h}}
            \newcommand{\hb}{{\cal i}}
            \newcommand{\eg}{{\em e.g.\ }}
            \newcommand{\ie}{{\em i.e.\ }}
            \newcommand{\ga}{{\gamma}}
            \newcommand{\la}{{\lambda}}
            
            \newcommand{\del}{{\delta}}
            \newcommand{\dal}{\dot{\alpha}}
            \newcommand{\dbeta}{\dot{\beta}}
            
            \newcommand{\dx}{\dot{x}}
            
            \newcommand{\dtheta}{\dot{\theta}}
            \newcommand{\baD}{\bar{D}}
            \newcommand{\bad}{\bar{d}}

            \newcommand{\batheta}{\bar{\theta}}
            
            \newcommand{\bapi}{\bar{\pi}}
            
            \newcommand{\bala}{\bar{\lambda}}
            \newcommand{\ra}{{\rightarrow}}
            
            \newcommand{\Lra}{{\Leftrightarrow}}
            \newcommand{\pet}{{\cal P}}

            \newcommand{\beq}{\begin{quote}}
            \newcommand{\eq}{\end{quote}}

            \newcommand{\al}{\alpha}
            \newcommand{\halv}{\frac{1}{2}}
            \newcommand{\ben}{\begin{enumerate}}
            \newcommand{\een}{\end{enumerate}}
            \newcommand{\bit}{\begin{itemize}}
            \newcommand{\ei}{\end{itemize}}

    	    \newcommand{\kvart}{{\frac{1}{4}}}
            
            \newcommand{\nn}{\nonumber}
     
            \newcommand{\r}[1]{(\ref{e:#1})}
            \newcommand{\edfl}[1]{\label{#1}\end{df}}
            
\begin{document}
            \begin{titlepage}
            \newpage
            \noindent
            ITP 94-14\\
            May 1994\\
	hep-th/9405164

            \vspace*{17 mm}
            \begin{center}{\LARGE\bf Superparticle  actions from
    superfields}\end{center}
            \vspace*{15 mm}
        \begin{center}{\large \bf R. Marnelius} and {\large \bf Sh. M.
            Shvartsman}
\footnote{Case Western Reserve University, Dept.
            of Phys., Cleveland, OH 44106, USA }\\
          \vspace*{10 mm} {\sl
            Institute of Theoretical Physics\\ Chalmers University of
            Technology\\ S-412 96  G\"{o}teborg, Sweden}\end{center}
            \vspace*{27 mm}
            \begin{abstract}
            Gauge invariant complex covariant actions for
            superparticles are derived from the field equations for
            the chiral superfields in a precise manner. The massive and
 massless cases  in four
            dimensions are treated both free and in interaction with an
            external super Maxwell field. By means of a generalized BRST
    quantization these complex actions are related to real actions with
    second class constraints  which are new in some cases. \end{abstract}
\end{titlepage}
            \newpage
            \setcounter{page}{1}
            \section{Introduction}
        There is now an enormous amount of literature on the problem of how to
 quantize superparticles
    in a manifestly supersymmetric and Lorentz
    covariant way. This problem has attracted such an interest  mainly since i
ts
    solution  is considered to
        be a crucial first step to a successful
        solution of the corresponding problem for superstrings.
        The problem is nontrivial mainly due to the appearance of second class
        constraints in the manifestly covariant actions.
        Such theories are not gauge theories and are therefore not possible to
    quantize in a gauge theoretical manner. The standard approach to circumven
t
        the problem is to modify the theories by means of auxiliary
        variables in such a way that the second class constraints
        are transformed into first class ones. The modified
        theories are then gauge theories which may be quantized by means
of standard BRST techniques.
 (We refrain from giving any references here since the literature is too vast.)

        Now there is also another BRST approach to the covariant
        quantization of theories with second class constraints which
        does not require the introduction of additional variables
\cite{RM} and which has not been applied
    to superparticles before. We shall apply this approach
and show that it to a large extent may be
    interpreted as an approach in which theories with second
        class constraints may be described by gauge invariant complex actions
        without second class constraints. Complex
        actions for gauge theories are not necessarily a bad feature since
 only
   the underlying physical theory is required to be implicitly decribed by a
 real action. However, it is true that for bosonic theories convergence
of the path integrals in
        general severely restricts the ranges of the bosonic variables
in the imaginary terms \cite{BG}. On
        the other hand no such restriction is necessary if the imaginary
terms involve fermionic variables
        \cite{MP}. For the complex superparticle models to be derived
here there are no such problems with
    the path integrals. For those who dislike the use of complex actions
it is in general possible to
    reformulate such models in terms of a real action with second class
constraints and apply the
    generalized BRST quantization of \cite{RM} without using the
 interpretation in terms of complex
actions.

        In distinction to previous approaches to the quantization of
superparticles our main concern in
    this paper is the precise connection between superfield equations and
the corresponding
    pseudoclassical particle models. This problem has attracted rather
little interest \cite{FSh,MSh,GLN}
    since it has no direct bearing on the quantization of the D=10
massless superparticle. However, it
    is relevant at a deeper level for our understanding of the relation
between particle models and field
    theory.  In this paper we shall apply the general method given
\cite{MM} which can be used as  a
    precise method to derive  particle models from quantum
        mechanical field equations. (In \cite{MM} it was used as a
general method to construct
    spinning particle models.) This method always yield gauge invariant
 particle actions which are not
    always real. However, when one obtains a complex action we shall
 demonstrate that there in general
    exists a corresponding real action usually then with second class
 constraints. The method is
    described below.

    In \cite{MM} it was shown that the corresponding particle model to a
 given set of field equations
    are described by a (pseudoclassical) Lagrangian whose phase space form is
            \be
            &&L=L_0+\la_i\phi_i
            \eel{e:01}
      where $L_0$ determines the class of fields, $\psi$, that is involved,
 and
        where
        $\phi_i$ are constraint variables and  $\la_i$  Lagrange multipliers.
        The constraint variables $\phi_i$ are
        classical expressions for the corresponding
        differential operators involved in the field equations, \ie
          \be
            &&\hat{\phi}_i\psi=0,\;\;\;i=1,\ldots
            \eel{e:02}
    Consistency requires the operators $\hat{\phi}_i$ to satisfy
    \be
    &&[\hat{\phi}_k, \hat{\phi}_l]_{\pm}=i\hat{F}_{klm}\hat{\phi}_m
    \eel{e:003}
    where $\hat{F}_{klm}$ are structure operators. Therefore, one may always
 find corresponding classical constraint
    variables, $\hat{\phi}_i\ra\phi_i$, that are of first class, \ie
satisfying the Poisson algebra
    \be
    &&\{{\phi}_i, {\phi}_j\}=F_{ijk} {\phi}_k
    \eel{e:004}
    where $F_{ijk}$ are structure functions. In other words
     \r{01} always represents a gauge invariant
        Lagrangian. However, since $\hat{\phi}_i$ are not always hermitian
      and since often not both $\hat{\phi}_i$
  and $\hat{\phi}^{\dag}_i$ are involved in the set of field equations \r{02},
        the Lagrangian \r{01} is in general complex since there may be complex
 constraint variables ${\phi}_i$
    for which $\bar{\phi}_i$ is not involved in \r{01}. As will be shown in
 section 4 the corresponding
    real Lagrangian is obtained by adding the independent complex conjugate
 constraints to $L$ which
    then  in general is no longer gauge invariant.

        In this paper we apply the above method to the field equations of a
 massive chiral scalar
        superfield in four dimensions  both free and in interaction with an
    external super Maxwell
        field. In all cases  the chiral superparticle
        is shown to be described by a complex covariant action which is gauge
 invariant. (Some of
these results  were presented  in
            \cite{MSh}.)
    The corresponding propagators which are obtained by means of
 the natural gauge theoretic  expressions in terms of the derived complex
        actions  are then shown to agree with the results of \cite{FSh}.
We also perform a corresponding treatment of the massless limits of the
field equations.
 The paper is organized as
follows: In section 2 we give  the correspondence between a particle
 Lagrangian and the
    class of scalar superfields. The free chiral scalar superfield case is
 then treated in section 3.
    Gauge invariant complex superparticle actions are derived. In section 4
we show that the
    quantization of these complex actions are related to a generalized BRST
 quantization of the
    conventional real Casalbuoni or Brink-Schwarz actions with second class
 constraints for free
    superparticles. Interaction with an external super Maxwell field by means
 of supersymmetric minimal
  coupling is considered in section 5. It is shown to be nontrivial to relate
 this construction to
 the corresponding superfield theory.
In section 6 we derive gauge invariant complex
    actions from the superfield equations in the interaction case. The
most natural corresponding real Lagrangians are shown to be equivalent to a
 Lagrangian obtained by
minimal coupling. In section 7 we derive gauge invariant complex
    covariant actions from the superfield equations in the massless limit. In
 this case interaction is
   trivially consistent with minimal coupling since this is also valid in the
 field equations. Some
final remarks are given in section 8.
 In an appendix we give some proofs for section 6.

    \setcounter{equation}{0}
        \section{Scalar superfields}
        Consider the class of superfields described by
        $\Phi(x,\theta,\batheta)$ where $x^{\mu}$ is the Minkowski coordinate
        in four dimensions (we are using spacelike metric \
        diag $\eta^{\mu\nu} = (-,+,+,+)$) and where $\theta^{\al}$ and
        $\batheta^{\dal}$ are odd Grassmann
        variables.
        The indices $\al$ and $\dal$ are two-spinor ones (we are using the
        notation of Ref. \cite{WB}).
        Such superfields may be viewed as wave functions in a quantum theory
        obtained from the quantization of the Lagrangian
            \be
            &L_0=p_{\mu}\dot{x}^{\mu}-i\pi^{\al}\dtheta_{\al}+i\bapi_{\dal}
            {\dot{\batheta^{\dal}}}
            \eel{e:5}
        where $p_{\mu}$ and
     $p_{\al}=i\pi_{\al}$, $\bar{p}_{\dal}=i\bapi_{\dal}$
    are conjugate momenta to $x^{\mu}$  and
        $\theta^{\al}$, $\batheta^{\dal}$ respectively.  This Lagrangian
        is real and leads
        to the hermitian operators $\hat{p}_{\mu}$ and $\hat{x}^{\mu}$
         satisfying
            \be
            &[ \hat{x}^{\mu}, \hat{p}_{\nu}]_-=i\del_{\nu}^{\mu}
            \eel{e:6}
        and to the odd fermionic operators $\hat{\pi}_{\al}, \hat{\theta}^
{\al},
        \hat{\bapi}_{\dal}$ and $\hat{\batheta}^{\dal}$
         satisfying
            \be
     &\hat{\bapi}_{\dal}=(\hat{\pi}_{\al})^{\dag},\;\;\;\hat{\batheta}^{\dal}=
         (\hat{\theta}^{\al})^{\dag}\nn\\
        &[\hat{\theta}^{\al}, \hat{\pi}_{\beta}]_+=\del_{\beta}^{\al},\;\;\;
         [\hat{\batheta}^{\dal}, \hat{\bapi}_{\dbeta}]_+=\del_{\dbeta}^{\dal}
            \eel{e:7}
        The corresponding nondegenerate state space is spanned by
 eigenstates to $\hat{x}^{\mu}, \hat{\theta}^{\al}$ and $\hat{\batheta}^{\dal}
$.
        We define the eigenstates $|x,\theta,\batheta\hb$ by
           \be
        &\hat{x}^{\mu}|x,\theta,\batheta\hb=x^{\mu}|x,\theta,\batheta\hb\nn\\
        &{\hat{\theta}}^{\al}|x,\theta,\batheta\hb=\theta^{\al}|x,\theta,
\batheta\hb\nn\\
        &\hat{\batheta}^{\dal}|x,\theta,\batheta\hb=\batheta^{\dal}|x,\theta
,\batheta\hb
            \eel{e:8}
            with the normalization
            \be
            &\vb
            x',\theta',\batheta'|x,\theta,\batheta\hb=\del^4(x-x')
            \del^2(\theta-\theta')\del^2(\batheta-\batheta')
            \eel{e:9}
        Any state $|\Phi\hb$ may then be expanded in terms of
        these eigenstates
            \be
            &&|\Phi\hb=\int\! d^4xd^2\theta d^2\batheta\,
            |x,\theta,\batheta\hb\Phi(x,\theta,\batheta)
            \eel{e:10}
        where the superfield $\Phi(x,\theta,\batheta)$ is given by
            \be
            &\Phi(x,\theta,\batheta)\equiv\vb
            x,\theta,\batheta|\Phi\hb
            \eel{e:11}

    \setcounter{equation}{0}
            \section{Gauge invariant complex actions for free superparticles}

        Our starting point is the equations for a free chiral
        scalar superfield in four dimensions: (see \eg eq.(9.24) in \cite{WB})
            \be
            &&\baD^2\bar{\Phi}(x,\theta,\batheta)=4m\Phi(x,\theta,\batheta)
\nn\\
            &&D^2\Phi(x,\theta,\batheta)=4m\bar{\Phi}(x,\theta,\batheta)
            \eel{e:03}
            where
            \be
     &&D^2\equiv D^{\al}D_{\al},\;\;\;\baD^2\equiv\baD_{\dal}\baD^{\dal}\nn\\
            &&D_{\al}\equiv
            \partial_{\al}+i\sigma^{\mu}_{\al\dbeta}\batheta^{\dbeta}
            \partial_{\mu}\nn\\
            &&\baD_{\dal}\equiv
            -\partial_{\dal}-i\theta^{\beta}\sigma^{\mu}_{\beta\dal}
            \partial_{\mu}
            \eel{e:04}
        Provided $m\neq0$ these equations imply
            \be
            &&D_{\al}\bar{\Phi}(x,\theta,\batheta)=0\nn\\
            &&\baD_{\dal}\Phi(x,\theta,\batheta)=0
            \eel{e:1}
    These chiral conditions are also used when \r{03} is derived from a
 Lagrangian density.
        Equ. \r{03} may be diagonalized:
            \be
            &&(\baD^2D^2-16m^2)\Phi(x,\theta,\batheta)=0\nn\\
            &&(D^2\baD^2-16m^2)\bar{\Phi}(x,\theta,\batheta)=0
            \eel{e:2}
        which again imply \r{1}. If we consider \r{1} as independent
        equations then \r{2} may be simplified to
            \be
            &(\partial^2 - m^2)\Phi(x,\theta,\batheta)=0\nn\\
            &(\partial^2 - m^2)\bar{\Phi}(x,\theta,\batheta)=0
            \eel{e:3}
        Thus, \r{1} and \r{3} are equivalent to \r{03}. Within the operator
        formulation the chiral superfield equations are then given by
            \be
            &(\hat{p}^2+m^2)|\Phi\hb=0\nn\\
            &(\hat{\bapi}_{\dal}-\hat{\theta}^{\beta}(\hat{p}\cdot\sigma)_
            {\beta\dal})|\Phi\hb=0
            \eel{e:4}
            Similarly the antichiral superfields equations may be written
            \be
            &(\hat{p}^2+m^2)|\bar{\Phi}\hb=0\nn\\
            &(\hat{\pi}_{\al}-(\hat{p}\cdot\sigma)_{\al\dbeta}\hat{\batheta}^
             {\dbeta})|\bar{\Phi}\hb=0
            \eel{e:12}
        where $|\bar{\Phi}\hb$ is related to $|\Phi\hb$ through the equation
            \be
            &&\bar{\Phi}(x,\theta,\batheta)=\overline{\vb\Phi|x, \theta,
\batheta\hb}=\vb x, \theta,
            \batheta|\bar{\Phi}\hb
            \eel{e:121}
 According to the method of ref.\cite{MM} the equations \r{4} may be considered
            to be obtained from the pseudoclassical Lagrangian
            \be
            &&L=L_0-\frac{e}{2}\phi+i\bala_{\dal}\bad^{\dal}
            \eel{e:13}
            where
            \be
            &\phi=p^2+m^2\nonumber\\
            &\bad_{\dal}=\bapi_{\dal}-\theta^{\beta}(p\cdot\sigma)_{\beta\dal}
            \eel{e:131}
           $L_0$ is given by \r{5} and  $e$ and $\bala^{\dal}$
            are Lagrange multipliers. Similarly the equations \r{12}
            lead to
            \be
            &&\bar{L}=L_0-\frac{e}{2}\phi-i\la^{\al}d_{\al}
            \eel{e:14}
            where
            \be
            &d_{\al}=\pi_{\al}-(p\cdot\sigma)_{\al\dbeta}\batheta^{\dbeta}
            \eel{e:141}
        The positive feature of these Lagrangians compared to the
        ones in refs.\cite{Ca,BS} is that they are both gauge
        invariant and manifestly supersymmetric and Lorentz
        invariant. The negative feature is that they are complex.
        Naively one would immediately discard complex actions.
        However, as we mentioned in the introduction complex actions for gauge
        theories is an allowed possibility and we shall also show that the
        above Lagrangians can be  used in a consistent fashion.

        Configuration space Lagrangians corresponding to \r{13} and
        \r{14} are easily obtained. A variation of $p^{\mu}$,
        $\pi_{\al}\;(\bapi_{\dal})$  and $\bala^{\dal}\;(\la^{\al})$ in
$L\;(\bar{L})$ yields
        equations which determine these variables. We find
            \be
            &&p^{\mu}=\frac{1}{e}(\dot{x}^{\mu}-
            i\theta^{\al}\sigma^{\mu}_{\al\dbeta}\bala^{\dbeta}),
            \;\;\;\bala^{\dal}=-\dot{\batheta}^{\dal},\;\;\;
            \bapi_{\dal}=\theta^{\al}(\sigma\cdot p)_{\al\dal}
            \eel{e:15}
        from $L$ and
            \be
            &&p^{\mu}=\frac{1}{e}(\dot{x}^{\mu}+
            i\la^{\al}\sigma^{\mu}_{\al\dbeta}\batheta^{\dbeta}),
            \;\;\;\la^{\al}=-\dot{\theta}^{\al},\;\;\;
            \pi_{\al}=(\sigma\cdot p)_{\al\dal}\batheta^{\dal}
            \eel{e:151}
   from $\bar{L}$. When these equations are inserted back into $L$ and
$\bar{L}$
        we get
            \be
            &&L_{conf}=\frac{(\dot{x}+i\theta\sigma\dot{\batheta})^2}{2e}-
            \frac{1}{2}em^2-i\pi_{\al}\dot{\theta^{\al}}\nn\\
&&\bar{L}_{conf}=\frac{(\dot{x}-i\dot{\theta}\sigma\batheta)^2}{2e}-
            \frac{1}{2}em^2+i\bapi_{\dal}\dot{\batheta^{\dal}}
            \eel{e:16}
        These Lagrangians are gauge invariant under the following gauge
transformations (the nonzero ones)
        \be
        &&\del e=\dot{\zeta},\;\;\;\del x^{\mu}=\zeta p^{\mu}
        \eel{e:162}
        where $\zeta$ is a real infinitesimal parameter function.
    The Lagrangian $L_{conf}$ is gauge invariant under the gauge
    transformations
        \be
        &&\del\batheta^{\dal}=i\bar{\kappa}^{\dal},\;\;\;\del
        x^{\mu}=\theta^{\al}\sigma^{\mu}_{\al\dal}\bar{\kappa}^{\dal},\;\;\;
        \del\pi_{\al}=-i(p\cdot\sigma)_{\al\dal}\bar{\kappa}^{\dal}
        \eel{e:163}
        where $\bar{\kappa}^{\dal}$ is an odd complex parameter function
    with $p^{\mu}$ given
    by \r{15}. The Lagrangian $\bar{L}_{conf}$ is invariant under
    \r{162} and
    the complex
    conjugation of \r{163} with $p^{\mu}$ given by \r{151}. It is a peculiar
property of the complex
    Lagrangians that we in the equations of motion \r{15},\r{151} and in the
 gauge transformations
    \r{163} must give up the strict reality properties of the involved
dynamical variables.

            The Lagrangians \r{16} should be compared with the
            Lagrangian given by Brink and Schwarz \cite{BS}
            \be
       &&L_{BS}=\frac{(\dot{x}-i\dot{\bar{\psi}}\ga\psi)^2}{2e}-\frac{1}{2}em^2
            \eel{e:161}
            where
            $\psi=\left(\begin{array}{c}\theta_{\al}\\\batheta^{\dal}
\end{array}
            \right)$  ($\bar{\psi}=(\theta^{\al},
    \batheta_{\dal})$) is a Majorana spinor and $\ga^{\mu}$ the Dirac matrices
        in the Weyl representation
        ($\dot{\bar{\psi}}\ga\psi\equiv\dot{\theta}\sigma\bar{\theta}-\theta
\sigma\dot{\bar{\theta}}$). Or
        equivalently  Casalbuoni's  Lagrangian \cite{Ca}
            \be
         &L_C=m[(\dot{x}-i\dot{\bar{\psi}}\ga\psi)^2]^{\frac{1}{2}} \eel{e:17}
obtained after eliminating the einbein variable $e$ when $m\neq 0$ in $L_{BS}$.
            Eliminating
            $\pi_{\al}\;(\bapi_{\dal})$ in $L_{conf}\;(\bar{L}_{conf})$ we
            find the following correspondence:
             \be
            &L\Lra L_{BS}(L_C)\; \mbox{ together with the constraint }
            \;\dot{\theta^{\al}}=0\nn\\
            &\bar{L}\Lra L_{BS}(L_C)\; \mbox{ together with the constraint
            }\; \dot{\batheta^{\dal}}=0
            \eel{e:18}

        (These external constraints may also be implemented by means of
 Lagrange multi\-pliers (like $\pi_{\al}(\bapi_{\dal})$ in \r{16}).) We have
 \eg the formal path integral relations\\
      \be &&\int\exp(i\int_0^sd\!\tau L)DpDxD\theta D\pi D\batheta D\bapi
            DeD\bala=\int\exp(i\int_0^sd\!\tau
            L_{BS})\del^2(\dot{\theta})DxD\theta D\batheta \nn\\
           &&\int\exp(i\int_0^sd\!\tau \bar{L})DpDxD\theta D\pi D\batheta D
\bapi
            DeD\la=
            \int\exp(i\int_0^sd\!\tau L_{BS})\del^2(\dot{\batheta})DxD\theta
            D\batheta\nn\\
            \eel{e:19}
For a possible definition of such path integrals over both bosonic and
 fermionic is \eg given in
section 2.4 and 6.4 of ref.\cite{FGS} where representations are given which
 automatically takes care
of the boundary conditions for the integration variables.
 Propagators in the proper time method are obtained by means of a natural gauge
        fixing of the Lagrange multipliers $e$ and $\bala^{\dal}\;(\la^{\al})$
        in the complex actions. The propagators for chiral and
        antichiral superfields are then given by
            \be
            &&\left.\vb\tau=s|\tau=0\hb\right|_{chiral}=\nn\\
            &&=\int\exp(i\int_0^sd\!\tau
            L)\del(e-1)\del^2(\bala)DpDxD\theta D\pi D\batheta D\bapi
            DeD\bala\nn\\
&&\left.\vb\tau=s|\tau=0\hb\right|_{antichiral}=\nn\\
            &&=\int\exp(i\int_0^sd\!\tau
            \bar{L})\del(e-1)\del^2(\la)DpDxD\theta D\pi D\batheta D\bapi
            DeD\la
             \eel{e:20}
            respectively. Now they are equal since
             \be
            &&\left.\vb\tau=s|\tau=0\hb\right|_{chiral}=\nn\\
            &&=\int\exp(i\int_0^sd\!\tau
         \frac{1}{2}((\dot{x})^2-m^2))\del^2(\dot{\theta})\del^2(\dot
{\batheta})
            DxD\theta D\batheta=\nn\\
            &&\left.\vb\tau=s|\tau=0\hb\right|_{antichiral}
            \eel{e:200}
            which is in agreement with the result of ref.\cite{FSh}. The
 results of this section were
    also given in \cite{MSh}.

    \setcounter{equation}{0}
    \section{Generalized BRST quantization of theories with second class
 constraints.}

    In order to give a brief description of the generalized BRST quantization
of theories with second
    class constraints given in \cite{RM} we consider a real Lagrangian of the
 form \r{01}
    \be
    && L=L_0+\la^i\phi_i
    \eel{e:4001}
    where $\la^i$ are Lagrange multipliers and $\phi_i$ constraint variables
 some of which  are
    considered to be of second class in Dirac's classification, \ie they
 satisfy PB-relations of the type
    \be
    &&\{\phi_i, \phi_j\}={U_{ij}}^k\phi_k+f_{ij}
    \eel{e:4002}
    where $f_{ij}\neq0$. According to \cite{RM} \r{4001} may be quantized by
 means of a generalized BRST
    quantization in which the BRST operator is  not nilpotent.  In the special
 case when
    ${U_{ij}}^k$ or $f_{ij}$ commutes with $\phi_i$  it is given by
    \be
    &&Q=\hat{\phi}_ic_i-\halv i {U_{ij}}^k\pet_kc^ic^j-\halv i {U_{ij}}^jc^i
    \eel{e:4003}
    where $c^i$ and $\pet_i$ are the ghosts and their conjugate momenta and it
 satisfies
    \be
    &&Q^2=\halv f_{ij}c^ic^j\neq 0
    \eel{e:4004}
    Provided it is conserved it may still be used to project out physical
 states by means of the BRST
    condition \be
    &&Q|ph\hb=0
    \eel{e:4005}
    In \cite{RMA} it was shown that in order for this condition to project out
 the appropriate states
    there must exist a bigrading such that (For the standard nilpotent case
 this was also obtained in
    \cite{HM}.)
     \be
    &&Q=\del+d
    \eel{e:4006}
    where
    \be
    &&\del^2=d^2=0,\;\;\;[\del, d]_+=Q^2
    \eel{e:4007}
    and such that the physical states are determined by
    \be
    &&\del|ph\hb=d|ph\hb=0
    \eel{e:4008}
    At this point we may connect this generalized BRST quantization with the
 method \cite{MM} since
    $\del$ (or $d$) could be interpreted as a complex BRST charge operator
 coming from a complex gauge
    invariant Lagrangian. If we in addition have
    \be
    &&d=\del^{\dag}
    \eel{e:4009}
which is necessary when the considered state space is  an inner product space,
 then $\del^{\dag}$
may
    be viewed as the BRST charge coming from the gauge invariant complex
 conjugate Lagrangian to the one
    for $\del$ and \r{4008} will then contain the solutions from both of
 these Lagrangians. Eq.\r{4008}
    may then also be interpreted such that a BRST quantization of a complex
 Lagrangian requires the use of
    two BRST operators; the natural non-hermitian one and its hermitian
 conjugate. Below we illustrate
    this procedure for the free superparticle.\\
\newpage

    {\bf The free superparticle.}

    The Brink-Schwarz Lagrangian \r{161} (or the Casalbuoni one \r{17} in the
 massive case) may be written
    as
    \be
    &&L_{BS}=L_0-\frac{e}{2}\phi-i\la^{\al}d_{\al}+i\bar{\la}_{\dal}\bar{d}^
{\dal}
    \eel{e:4010}
    in phase space where $L_0$ is given by \r{5}. This is the corresponding
real Lagrangian to \r{13} and
    \r{14} in which we have included all independent constraints. Notice that
 $d_{\al}$ and
    $\bar{d}_{\dal}$ are second class constraints ($\{d_{\al},
    \bar{d}_{\dal}\}=2i(p\cdot\sigma)_{\al\dal}$). According to the
generalized BRST procedure above
    we should then have the BRST charge operator
    \be
    &&Q=\eta\hat{\phi}+c^{\al}\hat{d}_{\al}-\bar{c}_{\dal}\hat{\bar{d}}^{\dal}
    \eel{e:4011}
    where $\hat{\phi},\;\hat{d}_{\al},\;\hat{\bar{d}}^{\dal}$ are
 corresponding operators to
    $\phi,\;d_{\al},\;\bar{d}^{\dal}$, and where $\eta$ is a hermitian
fermionic ghost and $c^{\al},\;\bar{c}_{\dal}$
    bosonic ghosts. $Q$ is hermitian and  assumed to be conserved. It
 satisfies \be
    &&Q=\del+\del^{\dag},\;\;\;\del^2=(\del^{\dag})^2=0,\;\;\;[\del,
    \del^{\dag}]_+=Q^2=-2c^{\al}\bar{c}^{\dal}(\hat{p}\cdot\sigma)_{\al\dal}
    \eel{e:4012}
    where
    \be
    &&\del=\halv\eta\hat{\phi}-\bar{c}_{\dal}\hat{\bar{d}}^{\dal},\;\;
    \;\del^{\dag}=\halv\eta\hat{\phi}+c^{\al}\hat{d}_{\al}
    \eel{e:4013}
    where $\del$ may be viewed as  the BRST charge from $L$ in \r{13},
and  $\del^{\dag}$ the one from
    $\bar{L}$ in \r{14}. Consider now the projection
    \be
    &&\del|ph\hb=\del^{\dag}|ph\hb=0
    \eel{e:4014}
    The chiral and antichiral superfield equations correspond
then to two different
    sectors or more precisely to two different choices for the
original state space in view of the
    treatment of ref. \cite{HM}. The chiral sector is obtained
by means of  the ghost fixing
    \be
    &&\pet|ph\hb=\bar{k}_{\dal}|ph\hb=0
    \eel{e:4015}
    where $\pet$ and $\bar{k}_{\dal}$ are canonical conjugate
momenta to the ghosts $\eta$ and
     $\bar{c}^{\dal}$ respectively. Consistency requires \cite{Aux}
    \be
    &&[Q, \pet]_+|ph\hb=[Q, \bar{k}_{\dal}]_-|ph\hb=0
    \eel{e:4016}
    or equivalently
    \be
    &&\hat{\phi}|ph\hb=\hat{\bar{d}}_{\dal}|ph\hb=0
    \eel{e:4017}
    and
    \be
    &&[Q^2, \pet]_{-}|ph\hb=[Q^2, \bar{k}_{\dal}]_{-}|ph\hb=0
    \eel{e:4018}
    which in the massive case implies
    \be
    &&c^{\al}|ph\hb=0
    \eel{e:4019}
    The conditions \r{4015} and \r{4019} fix completely the ghost dependence
of the physical states and
    \r{4017} are exactly the equations \r{4} for the chiral superfield.
Notice that \r{4015}, \r{4017}
    and \r{4019} imply \r{4014}.  In the massless limit \r{4018} does not
imply \r{4019} and we have
    ghost excitations from the point of view of ref. \cite{HM}.

    Similarly does
    \be
    &&\pet|ph\hb={k}_{\al}|ph\hb=0
    \eel{e:4020}
    in the massive case imply
    \be
    &&\bar{c}_{\dal}|ph\hb=0
    \eel{e:4021}
    and
    \be
    &&\hat{\phi}|ph\hb=\hat{{d}}_{\al}|ph\hb=0
    \eel{e:4022}
    which are exactly the equations \r{12} for the antichiral superfield. In
this case \r{4020}-\r{4022}
    imply \r{4014}.

    The above physical states are not inner product states. The BRST
quantization requires us therefore
    to work on bilinear forms of an original state space and its dual
 \cite{HM}.  If we make use of an
    extended BRST charge involving dynamical Lagrange multipliers and
antighosts then we may perform the
    BRST quantization on an inner product space from which we should obtain
 the propagators \r{200} in a
    precise manner (cf. \cite{Propa}). However, such a precise analysis will
 not be performed in this
    paper.

     \setcounter{equation}{0}
            \section{External fields introduced by minimal coupling}

    A natural way to introduce interactions with an external Maxwell
        superfield in a manifestly supersymmetric way is by means of the
 replacements
         \be
            &p_{\mu}\rightarrow p_{\mu}-g{A}_{\mu},\;\;\;
     d_{\al}\rightarrow
            d_{\al}+g{A_{\al}},\;\;\;\bad_{\dal}\rightarrow
            \bad_{\dal}+g\bar{A_{\dal}}
             \eel{e:25}
            in the constraints \r{131} and \r{141} in $L$ and $\bar{L}$
respectively. The super vector
    multiplet ($A_{\mu},\,A_{\al}, \bar{A}_{\dal}$) may be represented in
terms of a real scalar
    superfield $V=V(x,\theta,\batheta)$ as follows \cite{WB}
            \be
     &A_{\mu}\equiv \frac{1}{8}\bar{\sigma}_{\mu}^{\dal\al}[\baD_{\dal},
 D_{\al}]_-V,\;\;\;
     A_{\al}\equiv D_{\al}V,\;\;\;
            \bar{A}_{\dal}\equiv\baD_{\dal}V
            \eel{e:26}
         where
        $D_{\al}$ and $\baD_{\dal}$ are defined in \r{04}.
             The interaction Lagrangians  become then
            \be
            &&L'=L_0-\frac{e}{2}\phi'+i\bala_{\dal}\bad'^{\dal},\;\;\;
    \bar{L}'=L_0-\frac{e}{2}{\phi}'-i\la^{\al}d'_{\al}
          \eel{e:231}
         where $L_0$ is given by \r{5} and where
            \be
            &&\phi'\equiv (p-g{A})^2+m^2,\;\;\;
    d'_{\al}\equiv d_{\al}+gA_{\al},\;\;\;
            \bad'_{\dal}\equiv\bad_{\dal}+g\bar{A_{\dal}}
            \eel{e:271}

    According to the previous section  the corresponding real Lagrangian  to
 the complex conjugate
    ones in \r{231} should be
       \be
   &&L'_{real}=L_0-\frac{e}{2}\phi'+i\bala_{\dal}\bad'^{\dal}-i\la^{\al}
d'_{\al}
            \eel{e:274}
    which in configuration space becomes
    \be
    &&L'_{real,conf}=\frac{({\dot{x}}-i\dot{\bar{\psi}}\ga\psi)^2}{2e}+gA_
{\mu}({\dot{x}}^{\mu}-
    i\dot{\bar{\psi}}\ga^{\mu}\psi)
           +igA^{\al}\dot{\theta}_{\al}-ig\bar{A}_{\dal}\dot{\bar{\theta}}^
{\dal}-\frac{1}{2}em^2
            \eel{e:261}
    One may notice that this is a massive D=4 version of the massless D=10
Lagrangian
    given in \cite{RSNV}.
     A transition to configuration space for the complex Lagrangians \r{231}
yields on the other hand
             \be
            &&L'_{conf}=\frac{(\dot{x}+i\theta\sigma\dot{\batheta})^2}{2e}-
            \frac{1}{2}em^2-i\pi^{\al}\dot{\theta_{\al}}+gA\cdot(\dot{x}+i
\theta
            \sigma\dot{\batheta})
            -ig\bar{A}_{\dal}\dot{\batheta^{\dal}}\nn\\
           &&\bar{L}'_{conf}=\frac{(\dot{x}-i\dot{\theta}\sigma\batheta)^2}
{2e}-
            \frac{1}{2}em^2+i\bapi_{\dal}\dot{\batheta^{\dal}}+gA\cdot
(\dot{x}-i\dot{\theta}
            \sigma\batheta)+
            ig\dot{\theta}^{\al}A_{\al}
            \eel{e:28}
            from which we obtain  similar correspondences as in the free
            case
             \be
            &L'_{conf}\Lra L'_{real,conf}\; \mbox{ together with the
            constraint } \;\dot{\theta^{\al}}=0\nn\\
            &\bar{L}'_{conf}\Lra L'_{real,conf}\; \mbox{ together with the
            constraint }\; \dot{\batheta^{\dal}}=0
            \eel{e:29}

    Unfortunately this expected natural construction does not work. The
reason is that neither ($\phi',
    d'_{\al}$) nor ($\phi',
    \bar{d'}_{\dal}$) are first class constraints. We have
         \be
    &&\{d'_{\al}, d'_{\beta}\}=g\{A_{\al}, d_{\beta}\}+g\{d_{\al}, A_{\beta}\}
=ig[D_{\al},
    D_{\beta}]_+V=0\nn\\ &&\{\phi',
    d'_{\al}\}=2g(p^{\mu}-gA^{\mu})(-iD_{\al}A_{\mu}-\partial_{\mu}A_{\al})
            \eel{e:262}
    The only natural way to make them first class constraints is to require
 the last relation to be
    zero. In this case we would also have the same gauge invariance as in the
free case. However, this
    requires
            \be
            &A_{\mu}=i\partial_{\mu}V+{a}_{\mu}
            \eel{e:321}
            where ${a}_{\mu}$ is an antichiral field
            ($D_{\al}{a}_{\mu}=0$).  This
            severely restricts the superfield $V$. In fact it requires
            $\baD^2D_{\al}V=0$,
            a condition which is not allowed for the super
            Maxwell field (see \r{38} below).

            Thus, the constraints $d'_{\al}$ and  ${\phi}'$ are not  first
            class ones when $A^{\mu}$ is of the form \r{26}. This means that
the
        Lagrangians $L'$ and $\bar{L}'$ are not gauge invariant. In other
words,
            it is not
          possible to have a gauge invariant coupling to a general super vector
        multiplet when the latter is introduced by minimal coupling in the
free complex actions, and
there are of course no
    corresponding superfield equations. However, in the next section we shall
show that it is still
possible to obtain a real Lagrangian almost of the form \r{261}.

    \setcounter{equation}{0}
            \section{Gauge invariant introduction of external fields}

           In order to find the appropriate gauge invariant Lagrangians for
the superparticle in
    interaction with an external super Maxwell field we must start from the
corresponding field
    equations and apply the method of \cite{MM} described in the
introduction. Now a  real
    field Lagrangian for scalar chiral superfields in interactions
        with an external
        super Maxwell field seems to require at least a doublet of
superfields. Define therefore
        \be
            &&\bf\Phi\equiv\left(\begin{array}{c}\Phi_+\\ \Phi_-\end{array}
\right),\;\;\;
    \bar{\bf\Phi}\equiv\left(\begin{array}{c}\bar\Phi_+\\ \bar\Phi_-
\end{array}\right)
            \eel{e:33}
            The super field equations  from this Lagrangian is  (see
\eg ref.\cite{WB})
            \be
            &&\bar{D}^2e^{g\sigma_3V}\bar{\bf\Phi}=4m\sigma_1{\bf\Phi}\nn\\
            &&D^2e^{g\sigma_3V}{\bf\Phi}=4m\sigma_1\bar{\bf\Phi}
            \eel{e:34}
            where $V$ is a real external scalar superfield, $\sigma_i$ the
Pauli matrices, and
            $g$ as before a coupling constant. Even these equations imply the
chiral conditions
            \r{1} for $m\neq0$, in which case \r{34} may also be diagonalized
to yield
            \be
            &&\left(\bar{D}^2e^{-g\sigma_3V}D^2e^{g\sigma_3V}-16m^2\right)
{\bf\Phi}=0\nn\\
            &&\left(D^2e^{-g\sigma_3V}\bar{D}^2e^{g\sigma_3V}-16m^2\right)
\bar{\bf\Phi}=0
            \eel{e:35}
            In terms of the components $\Phi_{\pm}$ in \r{33} they become
            \be
            &&\left(\bar{D}^2e^{\mp gV}D^2e^{\pm gV}-16m^2\right)\Phi_{\pm}=0
\nn\\
            &&\left(D^2e^{\mp gV}\bar{D}^2e^{\pm gV}-16m^2\right)\bar{\Phi}_
{\pm}=0
            \eel{e:36}
            Even \r{35} or \r{36} implies the chiral conditions \r{1}. If we
as in the free case choose
            \r{1} as separate independent equations then \r{36} may be
simplified
            to
            \be
            &&\left\{(i\partial_{\mu}\pm gA'_{\mu})^2+m^2\mp\frac{g}{4}[D_
{\al},
            W^{\al}]_-+\frac{g^2}{2}A^{\al}W_{\al}\right\}\Phi_{\pm}=0\nn\\
            &&\left\{(i\partial_{\mu}\mp
            g\bar{A'}_{\mu})^2+m^2\pm\frac{g}{4}[\bar{D}_{\dal},
            \bar{W}^{\dal}]_-+\frac{g^2}{2}
            \bar{W}_{\dal}\bar{A}^{\dal}\right\}\bar{\Phi}_{\pm}=0
            \eel{e:37}
             where  $A_{\al}$ and  $\bar{A}_{\dal}$ are given
            by \r{26} , and where
            \be
    &&{ A}'_{\mu}\equiv A_{\mu}+\halv
    i\partial_{\mu}V=\kvart\bar{\sigma}^{\dal\al}_{\mu}\bar{D}_{\dal}D_
{\al}V,\nn\\
            &&\bar{ A}'_{\mu}=A_{\mu}-\halv
    i\partial_{\mu}V=-\kvart\bar{\sigma}^{\dal\al}_{\mu}D_{\al}\bar{D}_
{\dal}V,\nn\\
    &&W_{\al}\equiv-\frac{1}{4}\bar{D}^2D_{\al}V,\;\;\;
            \bar{W}_{\dal}\equiv -\frac{1}{4}D^2\bar{D}_{\dal}V
            \eel{e:38}
           where in turn $A_{\mu}$ is the real super vector field defined
in \r{26}. $W_{\al}$ and
    $\bar{W}_{\dal}$ are the gauge invariant supersymmetric field strengths.

    There is a peculiar feature of the use of chiral Lagrangians in this
case: Although $\bar{\bf\Phi}$
    is assumed to be the complex conjugate to ${\bf\Phi}$ in the Lagrangian
density this is no longer
    true in the equations of motion. However, from \r{37} it looks like we
may treat $\bar{\Phi}_{\mp}$
    as the complex conjugate to ${\Phi}_{\pm}$ since the operators in \r{37}
then are related by hermitian
    conjugation.  From the corresponding
            operator expressions of \r{37} and \r{1}
     ($-i\partial_{\mu}\ra p_{\mu}$,
            $D_{\al}\ra d_{\al}$ and $\bar{D}_{\dal}\ra -\bar{d}_{\dal}$) we
may therefore derive the
    following
        interaction Lagrangians for the
            scalar fields $\Phi_{\pm}$, $\bar{\Phi}_{\mp}$:
            \be
            &&L''_{\pm}=L_0-\frac{e}{2}\phi''_{\pm}
    +i\bala_{\dal} \bad^{\dal}\nn\\
            &&{\bar{L}''}_{\pm}=L_0-\frac{\bar{e}}{2}\bar{\phi}''_{\pm}
    -i\la^{\al} d_{\al}
            \eel{e:39}
            where $L_0$ is given by \r{5} and  $d^{\al}$ and $\bad^{\dal}$
are defined in \r{131}
    and \r{141} respectively, and where
            \be
            &&\phi''_{\pm}\equiv
            (p\mp g
            A')^2+m^2\pm\frac{g}{2}d^{\al}W_{\al}+\frac{g^2}{2}
A^{\al}W_{\al}\nn\\
            &&\bar{\phi}''_{\pm}\equiv
            (p\mp g\bar{A'})^2+m^2\pm\frac{g}{2}\bar{d}_{\dal}\bar{W}^{\dal}
            +\frac{g^2}{2}\bar{W}_{\dal}\bar{A}^{\dal}
     \eel{e:24}
Notice that $L''_{+}$ and $L''_{-}$ only differ in the signs
of the coupling constant $g$. In the appendix it is proved that
$\phi''_{\pm}$ and $\bar{d}_{\dal}$ satisfy the same
Poisson algebra as $\phi$ and $\bar{d}_{\dal}$ do in the free case.
The Lagrangians \r{39} are therefore gauge invariant under the
same abelian gauge group. $L''_{\pm}$ is invariant
    under (the nonzero transformations)
     \be
    &&\del x^{\mu}=\zeta(p^{\mu}\mp g{A'}^{\mu}\mp\frac{g}{4}W\sigma^{\mu}
\batheta),\;\;\;\del
    e=\dot{\zeta},\;\;\;\del\theta^{\al}=\pm i\frac{\zeta
    g}{4}W^{\al},\nn\\
    &&\del\pi_{\al}=-i\frac{\zeta}{2}\partial_{\al} \phi''_{\pm},\;\;\;
    \del\bapi_{\dal}=-i\frac{\zeta}{2}\partial_{\dal}
\phi''_{\pm},\;\;\;\del p_{\mu}=-\frac{\zeta}{2}\partial_{\mu} \phi''_{\pm}
\nn\\
    \eel{e:391}
    and
    \be
    &&\del x^{\mu}=\theta\sigma^{\mu}\bar{\kappa},\;\;\;\del
    \bar{\theta}^{\dal}=i\bar{\kappa}^{\dal},\nn\\
    &&\del
\pi_{\al}=-i(p\cdot\sigma)_{\al\dal}\bar{\kappa}^{\dal},\;\;\;\del\bala^{\dal}
=-i\dot{\bar{\kappa}^{\dal}}
    \eel{e:392}
     where $\zeta$ is a (real) infinitesimal parameter function in \r{391} and
   $\kappa^{\al}$ an odd complex parameter function in \r{392}.
 $\bar{L}''_{\pm}$ is invariant under the corresponding complex conjugated
transformations.

            The corresponding
    configuration space Lagrangians to \r{39} may \eg be written as
            \be
            &L''_{\pm
            ({conf})}=&\frac{1}{2e}(\dx -i\dot{\bar{\psi}}\ga\psi)^2
            \pm
            gA'\cdot(\dx -i\dot{\bar{\psi}}\ga\psi)-\frac{e}{2}m^2\pm
            igA^{\al}\dot{\theta}_{\al}+\nn\\
            &&-\pi^{\al}
            (i\dot{{\theta}}_{\al}\pm\frac{eg}{4}{W}_{\al})\nn\\
            &{\bar{L}''}_{\pm
            (conf)}=&\frac{1}{2\bar{e}}(\dx -i\dot{\bar{\psi}}\ga\psi)^2
            \pm
            g\bar{A'}\cdot(\dx -i\dot{\bar{\psi}}\ga\psi)-
            \frac{\bar{e}}{2}m^2\pm
            ig\bar{A}_{\dal}\dot{\batheta}^{\dal}+\nn\\
            &&+\bar{\pi}_{\dal}
            (i\dot{\bar{\theta}}^{\dal}\mp\frac{\bar{e}g}{4}\bar{W}^{\dal})
            \eel{e:40}
    Imposing $\dot{\theta}^{\al}=\pm i\frac{eg}{4}W^{\al}$
for $L''_{\pm}$ we have effectively  gauge
    invariance under ($\beta\equiv\zeta/e$)
     \be
    &&\del x^{\mu}=\beta(\dot{x}+i\theta\sigma^{\mu}
\dot{\bar{\theta}}),\;\;\;\del e=\dot{\beta}
    e+\dot{e}\beta,\;\;\;\del\theta^{\al}=\beta\dot{\theta}^{\al},\;\;\;\del
{\bar{\theta}}^{\dal}=0
    \eel{e:401}
    as well as and under the transformations on the first line in \r{392}.
For ${\bar{L}''}_{\pm(conf)}$
    we have invariance under the corresponding complex conjugated
transformations.
    Notice that these gauge transformations are
    the same as those in the free case except for the last one in \r{401}.

      If we use the same gauge fixing
    as in \r{20} we find the propagators
            \be
            &&\left.\vb\tau=s|\tau=0\hb_{(\pm)}\right|_{chiral}=\nn\\
            &&=\int\exp(i\int_0^sd\!\tau
            L''_{\pm})\del(e-1)\del^2(\bala)DpDxD\theta D\pi D\batheta D\bapi
            DeD\bala=\nn\\
            &&=\int\exp(i\int_0^sd\!\tau
        L'_{(\pm)})\del^2(\dot{\theta}\mp i\frac{g}{4}{W})\del^2(
\dot{\batheta})
            DxD\theta D\batheta\nn\\
            &&\left.\vb\tau=s|\tau=0\hb_{(\mp)}\right|_{antichiral}=\nn\\
            &&=\int\exp(i\int_0^sd\!\tau
   {\bar{L}''}_{\pm})\del(\bar{e}-1)\del^2(\la)DpDxD\theta D\pi D\batheta D
\bapi
            D\bar{e}D\la=\nn\\
            &&=\int\exp(i\int_0^sd\!\tau
            \bar{L}'_{(\pm)})\del^2(\dot{\theta})
    \del^2(\dot{\batheta}\pm i\frac{g}{4}\bar{W})
            DxD\theta D\batheta
             \eel{e:41}
            where
            \be
       &L'_{(\pm)}=&\frac{1}{2}(\dx -i\dot{\bar{\psi}}\ga\psi)^2-\frac{1}{2}m^2
            \pm g{ A'}\cdot(\dx -i\dot{\bar{\psi}}\ga\psi)\nn\\
            &&
           \pm igA^{\al}\dot{\theta}_{\al}\mp ig\bar{A}_{\dal}
    \dot{\batheta}^{\dal}
      \eel{e:42}
is equal to  $L'_{real,conf}$ in  \r{261} with $e=1$ and
$A_{\mu}$ replaced by ${A'}_{\mu}$ defined in \r{38}.

       The expressions \r{41} agree with those of \cite{FSh} except for an
interchange of the
indices $\pm$ in the antichiral case. Notice that we have used the Weyl
 ordering
    when we \eg replaced the commutator $[D_{\al}, W^{\al}]_-$ by
        $2d^{\al}W_{\al}$ in the constraints $\phi_{\pm}''$. Therefore the
Lagrangians \r{40} agrees up to
        the term $(D_{\al}W^{\al})$ in $L''_{\pm(conf)}$
(and $(\bar{D}_{\dal}\bar{W}^{\dal})$ in
        ${\bar{L}''}_{\pm
            (conf)}$) with those obtained in
        \cite{FSh}. (There the operators  in \r{37} also play a crucial role
although they were obtained
    in a different manner.)

    The BRST treatment in section 4 suggests that the corresponding real
Lagrangian to the complex
    conjugate pair \r{39} is
    \be
    &&L''_{\pm(real)}=L_0 -\frac{e}{2}\phi''_{\pm}
     -\frac{\bar{e}}{2}\bar{\phi}''_{\pm}+i\bala_{\dal} \bad^{\dal}
    -i\la^{\al} d_{\al}
    \eel{e:60}
    which in configuration space becomes
    \be
    &&L''_{\pm(real,conf)}=\halv\frac{({\dot{x}}^{\mu}-i\dot{\bar{\psi}}\ga^
{\mu}\psi)}{e_1}\pm
    gA_{\mu}({\dot{x}}^{\mu}- i\dot{\bar{\psi}}\ga^{\mu}\psi)-\frac{1}{2}e_1
m^2\nn\\
    &&\mp g\frac{e_2}{2e_1}\partial_{\mu}V({\dot{x}}^{\mu}-
i\dot{\bar{\psi}}\ga^{\mu}\psi)
    +\frac{g^2(e_1^2+e_2^2)}{8e_1}(\partial\cdot V)^2\nn\\
    &&
    -\frac{g^2}{8}e_1(A^{\al}W_{\al}+\bar{W}_{\dal}\bar{A}^{\dal})-
    i\frac{g^2}{8}e_2(A^{\al}W_{\al}-\bar{W}_{\dal}\bar{A}^{\dal})
    \eel{e:61}
    where $e_1\equiv e+\bar{e}$ and $e_2\equiv i(\bar{e}-e)$.

    It is rather obvious that $\bar{\bf\Phi}$ is not
    the complex conjugate superfield to $\bf\Phi$ in \r{34}. One may also
notice  that the free chiral
    superfield equations \r{1} are only obtained in the limit  $g\,\ra\, 0$
if $\bar\Phi_{\mp}$ is assumed
    to be complex conjugates to $\Phi_{\pm}$ as we did above. However, even
when this redefinition
    ($\sigma_1\bar{\bf\Phi}\,\ra\,\bar{\bf\Phi}$) is inserted into the
equations \r{34} we still do not
    obtain equations which are manifestly consistent with complex
conjugation. Such equations are only
    obtained if we introduce the following fields
    \be
    &&{\bf\Phi'}\equiv e^{\frac{g}{2}\sigma_3V}{\bf\Phi},\;\;\;
\bar{\bf\Phi}'\equiv
    e^{-\frac{g}{2}\sigma_3V}\sigma_1\bar{\bf\Phi}
    \eel{e:101}
    since \r{34} then
    becomes \be
    &&\bar{D'}^2\bar{\bf\Phi}'=4m{\bf\Phi'},\;\;\;{D'}^2{\bf\Phi'}=4m\bar{\bf
\Phi}'
    \eel{e:102}
    where
    \be
    &&D'_{\al}\equiv
    e^{-\frac{g}{2}\sigma_3V}D_{\al}e^{\frac{g}{2}\sigma_3V}=D_{\al}+
\frac{g}{2}\sigma_3A_{\al},\nn\\
    &&
    \bar{D'}_{\dal}\equiv
    e^{\frac{g}{2}\sigma_3V}\bar{D}_{\dal}e^{-\frac{g}{2}\sigma_3V}=
\bar{D}_{\dal}-\frac{g}{2}\sigma_3\bar{A}_{\dal}
    \eel{e:103}
    The fields ${\bf\Phi'},\,(\bar{\bf\Phi}')$ are no longer chiral ones
since \r{101} transforms the
    chiral conditions \r{1} into
  \be
    &&D'_{\al}\bar{\bf\Phi}'=0,\;\;\;\bar{D'}_{\dal}{\bf\Phi}'=0
    \eel{e:104}
    Notice that \r{101} and \r{104} are manifestly consistent with complex
conjugation.
    Equ. \r{102} may  be diagonalized to
            \be
            &&\left(\bar{D'}^2D'^2-16m^2\right){\bf\Phi'}=0\nn\\
            &&\left(D'^2\bar{D'}^2-16m^2\right)\bar{\bf\Phi'}=0
            \eel{e:105}
             If we like in the free case choose
        \r{104} as separate independent equations then \r{35} may be simplified
            to
            \be
     &&\left\{(i\partial_{\mu}\pm gA_{\mu})^2+m^2\mp\frac{g}{4}[D^{(\pm)}_
{\al},
            W^{\al}]_-\right\}\Phi'_{\pm}=0\nn\\
            &&\left\{(i\partial_{\mu}\pm
            g{A}_{\mu})^2+m^2\mp\frac{g}{4}[\bar{D}^{(\pm)}_{\dal},
            \bar{W}^{\dal}]_-\right\}\bar{\Phi}'_{\pm}=0
            \eel{e:106}
    in terms of the components of  ${\bf\Phi}'$ and ${\bar{\bf\Phi}}'$.
     Here  $A_{\mu}$
    is the real vector superfield defined in \r{26}
          and
            \be
    &&D^{(\pm)}_{\al}\equiv D_{\al}\pm\frac{g}{2}A_{\al},\;\;\;\bar{D}^
{(\pm)}_{\dal}\equiv
    \bar{D}_{\dal}\mp\frac{g}{2}\bar{A}_{\dal}
            \eel{e:107}
           These
     relations correspond to \r{103} above.  From the
            operator expressions in \r{104} and \r{106}
     (using $-i\partial_{\mu}\ra p_{\mu}$,
            $D_{\al}\ra d_{\al}$ and $\bar{D}_{\dal}\ra -\bar{d}_{\dal}$) we
find now the following
interaction Lagrangians for the scalar fields
$\Phi'_{\pm}$, $\bar{\Phi'}_{\pm}$:
            \be
            &&L'''_{\pm}=L_0-\frac{e}{2}\phi'''_{\pm}
    +i\bala_{\dal} \bad^{\dal}_{(\pm)}\nn\\
            &&{\bar{L}'''}_{\pm}=L_0-\frac{\bar{e}}{2}\bar{\phi}'''_{\pm}
    -i\la^{\al} d_{(\pm)\al}
            \eel{e:108}
            where $L_0$ is given by \r{5} and where
            \be
    && d_{(\pm)\al}\equiv  d_{\al}\pm\frac{g}{2}A_{\al},\;\;\;\bad^{\dal}_
{(\pm)}\equiv
    \bad^{\dal}\pm \frac{g}{2}\bar{A}^{\dal}\nn\\
            &&\phi'''_{\pm}\equiv
            (p\mp g
            A)^2+m^2\mp\frac{g}{2}d^{(\pm)}_{\al}W^{\al}\nn\\
            &&\bar{\phi}'''_{\pm}\equiv
            (p\mp g{A})^2+m^2\pm\frac{g}{2}\bar{d}^{(\pm)}_{\dal}\bar{W}^
{\dal}
     \eel{e:109}
            In the appendix it is proved that $\phi'''_{\pm}$ and $\bar{d}_
{\dal}^{(\pm)}$ satisfy the
    same Poisson algebra as $\phi$ and $\bar{d}_{\dal}$ in the free case. The
Lagrangians \r{108} are
    therefore gauge invariant under the same abelian gauge group. $L'''_{\pm}
$ is invariant
    under (the nonzero transformations)
     \be
    &&\del x^{\mu}=\zeta(p^{\mu}\mp g{A}^{\mu}\mp\frac{g}{4}W\sigma^{\mu}
\batheta),\;\;\;\del
    e=\dot{\zeta},\;\;\;\del\theta^{\al}=\pm i\frac{\zeta
    g}{4}W^{\al},\nn\\
    &&\del\pi_{\al}=-i\frac{\zeta}{2}\partial_{\al} \phi'''_{\pm},\;\;\;
    \del\bapi_{\dal}=-i\frac{\zeta}{2}\partial_{\dal}
    \phi'''_{\pm},\;\;\;\del p_{\mu}=-\frac{\zeta}{2}\partial_{\mu} \phi'''
_{\pm}
    \eel{e:110}
    and
    \be
    &&\del x^{\mu}=\theta\sigma^{\mu}\bar{\kappa},\;\;\;\del
    \bar{\theta}^{\dal}=i\bar{\kappa}^{\dal}, \;\;\;\del
    p_{\mu}=\mp\frac{g}{2}\bar{\kappa}^{\dal}\partial_{\mu}\bar{A}_{\dal}\
nn\\
     &&\del
    \pi_{\al}=-i(p\cdot\sigma)_{\al\dal}\bar{\kappa}^{\dal}\pm
    i\frac{g}{2}\bar{\kappa}^{\dal}\partial_{\al}\bar{A}_{\dal},\;\;\;\del
\bar{\pi}_{\dal}=\pm
    i\frac{g}{2}\bar{\kappa}^{\dot{\beta}}\partial_{\dal}\bar{A}_{\dot
{\beta}}\;\;\;\del\bala^{\dal}=-i\dot{\bar{\kappa}^{\dal}}
    \eel{e:111}  while  $\bar{L}'''_{\pm}$ is invariant under the complex
conjugated transformations.

            The corresponding
    configuration space Lagrangians to \r{108} may \eg be written as
            \be
            &L'''_{\pm
            ({conf})}=&\frac{1}{2e}(\dx -i\dot{\bar{\psi}}\ga\psi)^2
            \pm
            gA\cdot(\dx -i\dot{\bar{\psi}}\ga\psi)-\frac{e}{2}m^2\pm
            i\frac{g}{2}A^{\al}\dot{\theta}_{\al}\mp
            i\frac{g}{2}\bar{A}_{\dal}\dot{\batheta}^{\dal}-\nn\\
            &&-\pi^{\al}
            (i\dot{{\theta}}_{\al}\pm\frac{eg}{4}{W}_{\al})\nn\\
            &{\bar{L}'''}_{\pm
            (conf)}=&\frac{1}{2\bar{e}}(\dx -i\dot{\bar{\psi}}\ga\psi)^2
            \pm
            g{A}\cdot(\dx -i\dot{\bar{\psi}}\ga\psi)-
            \frac{\bar{e}}{2}m^2\pm
            i\frac{g}{2}A^{\al}\dot{\theta}_{\al}\mp
            i\frac{g}{2}\bar{A}_{\dal}\dot{\batheta}^{\dal}+\nn\\
            &&+\bar{\pi}_{\dal}
            (i\dot{\bar{\theta}}^{\dal}\mp\frac{\bar{e}g}{4}\bar{W}^{\dal})
            \eel{e:112}
    When  $\dot{\theta}^{\al}=\pm i\frac{eg}{4}W^{\al}$ is imposed on
$L'''_{\pm}$ we have effectively
    gauge invariance under \r{401}.

     The corresponding real Lagrangian to the complex
    conjugate pair \r{112} is given by
    \be
    &&L'''_{\pm(real)}=L_0 -\frac{e}{2}\phi'''_{\pm}
     -\frac{\bar{e}}{2}\bar{\phi}'''_{\pm}+i\bala_{\dal} \bad^{\dal}_{(\pm)}
    -i\la^{\al} d_{\al}^{(\pm)}
    \eel{e:113}
    which in configuration space becomes
    \be
    &&L'''_{\pm(real,conf)}=\halv\frac{({\dot{x}}-i\dot{\bar{\psi}}\ga\psi)^2}
{e_1}\pm
    gA_{\mu}({\dot{x}}^{\mu}- i\dot{\bar{\psi}}\ga^{\mu}\psi)-\frac{1}{2}e_1
m^2\nn\\
    &&\pm
            i\frac{g}{2}A^{\al}\dot{\theta}_{\al}\mp
            i\frac{g}{2}\bar{A}_{\dal}\dot{\batheta}^{\dal}
    \eel{e:114}
    where $e_1=e+\bar{e}$. This is obviously a much nicer Lagrangian than
\r{61} obtained before
    which indicates that $\bf\Phi'$ and $\bar{\bf\Phi}'$ are the true complex
conjugate pair of fields. It
    is very close to \r{261} obtained by minimal coupling.     In fact, we
get exact agreement if
 $g$ is replaced by $g/2$ in the last two replacements in
\r{25} since the resulting modified \r{261} is \r{114}.
This minimal coupling is also suggested by the relation
\be
&&\{d_{\al}^{(\pm)}, \bar{d}_{\dal}^{(\pm)}\}=2i(p_{\mu}\mp gA_{\mu})\sigma^
{\mu}_{\al\dal}
\eel{e:1141}
A comparison between \r{274} and \r{113} yields that the Lagrange multiplier
$\la^{\al}$ and
$\bar{\la}^{\dal}$ in \r{274} are equal to $\la^{\al}\mp\frac{eg}{2}W^{\al}$
and
$\bar{\la}^{\dal}\mp\frac{\bar{e}g}{2}\bar{W}^{\dal}$ respectively in terms
of the Lagrange
multipliers in \r{113}. This is an important difference since even when this
modified minimal coupling
is imposed on the free complex actions \r{16} it does
 not lead to gauge invariant complex covariant actions
and corresponds to no superfield equations.

      The propagators are found as before. Using the same gauge fixing
    as in \r{20} we find the following propagators for $\Phi'_{\pm}$
            \be
            &&\vb\tau=s|\tau=0\hb_{(\pm)}=\nn\\
            &&=\int\exp(i\int_0^sd\!\tau
            L'''_{\pm})\del(e-1)\del^2(\bala)DpDxD\theta D\pi D\batheta D\bapi
            DeD\bala=\nn\\
            &&=\int\exp(i\int_0^sd\!\tau
            L'''_{\pm (real)})\del^2(\dot{\theta}\mp i\frac{g}{4}{W})\del^2(
\dot{\batheta})
            DxD\theta D\batheta
             \eel{e:115}
    and for $\bar{\Phi}'_{\pm}$
    \be
            &&\vb\tau=s|\tau=0\hb_{(\pm)}=\nn\\
            &&=\int\exp(i\int_0^sd\!\tau
    \bar{L}'''_{\pm})\del(\bar{e}-1)\del^2(\la)DpDxD\theta D\pi D\batheta D
\bapi
            D\bar{e}D\bala=\nn\\
            &&=\int\exp(i\int_0^sd\!\tau
            L'''_{\pm (real)})\del^2(\dot{\batheta}\pm i
\frac{g}{4}{W})\del^2(\dot{\theta})
            DxD\theta D\batheta
             \eel{e:116}
            where
            $L'''_{\pm(real)}$ is
  equal to  $L'''_{\pm(real,conf)}$ in  \r{114} with $e_1=1$. These
expressions are in fact
    consistent with \r{41} since
    \be
    &&L'_{(\pm)}-L'''_{\pm(real)}=\pm\halv ig\frac{dV}{d\tau}
    \eel{e:117}
     inside the path integrals \r{41} and \r{115}.

 According to the generalized BRST procedure briefly described in section 4
 we should have the following hermitian BRST charge operators to
the Lagrangians \r{113}
    \be
    &&Q_{\pm}=\eta\hat{\phi}'''_{\pm}+\bar{\eta}\hat{\bar{\phi}}'''_{\pm}+
    c^{\al}\hat{d}_{\al}^{(\pm)}-\bar{c}_{\dal}\hat{\bar{d}}^{\dal}_{(\pm)}
    \eel{e:611}
 where  $\eta,\;\bar{\eta}$ are fermionic ghosts and $c^{\al},\;\bar{c}_{\dal}$
    bosonic spinor ghosts. $Q_{\pm}$ are assumed to be conserved.
Eq.\r{611} may be written
     \be
    &&Q_{\pm}=\del_{\pm}+{\del}^{\dag}_{\pm}
    \eel{e:612}
    where
    \be
    &&\del_{\pm}=\eta\hat{\phi}'''_{\pm}-\bar{c}_{\dal}
\hat{\bar{d}}^{\dal}_{(\pm)},\;\;
    \;{\del}^{\dag}_{\pm}=\bar{\eta}\hat{\bar{\phi}}'''_{\pm}+
c^{\al}\hat{d}_{\al}^{(\pm)}
    \eel{e:613}
    which satisfy
    \be
    &&\del^2_{\pm}=(\del_{\pm}^{\dag})^2=0,\;\;\;[\del_{\pm},
    \del_{\pm}^{\dag}]_+=Q_{\pm}^2=c^{\al}\bar{c}^{\dal}
[\hat{d}_{\al}^{(\pm)},\hat{\bar{d}}_{\dal}^{(\pm)}]_+\nn\\
    &&+\eta\bar{\eta}
    [\hat{\phi}'''_{\pm},\hat{\bar{\phi}}'''_{\pm}]_-+\eta
 c^{\al}[\hat{\phi}'''_{\pm},\hat{d}_{\al}^{(\pm)}]_--\bar{\eta}\bar{c}_{\dal}
    [\hat{\bar{\phi}}'''_{\pm},\hat{\bar{d}}^{\dal}_{(\pm)}]_- \eel{e:614}
    where
    \be
    &&[\hat{d}_{\al}^{(\pm)},\hat{\bar{d}}_{\dal}^{(\pm)}]_+\neq0,
\;\;\;
   [\hat{\phi}'''_{\pm},\hat{\bar{\phi}}'''_{\pm}]_-\neq
    0,\nn\\
    &&[\hat{\phi}'''_{\pm},\hat{d}_{\al}^{(\pm)}]_-\neq 0,
   \;\;\;[\hat{\bar{\phi}}'''_{\pm},\hat{\bar{d}}^{\dal}_{(\pm)}]\neq 0
    \eel{e:615}
     $\del_{\pm}$ and  $\del_{\pm}^{\dag}$ may be viewed as  the BRST charges
 from $L'''_{\pm}$ and
    $\bar{L}'''_{\pm}$ in \r{108}.

    Consider now the projections
    \be
    &&\del_{\pm}|ph\hb={\del}^{\dag}_{\pm}|ph\hb=0
    \eel{e:616}
    The genuine physical sectors may then be obtained by means of
    auxiliary conditions \cite{Aux}. The following ghost fixing is possible
to impose
    \be
    &&\pet|ph\hb=\bar{k}_{\dal}|ph\hb=0
    \eel{e:617}
    where $\pet$ and $\bar{k}_{\dal}$ are canonical conjugate momenta to the
ghosts $\eta$ and
     $\bar{c}^{\dal}$ respectively. Consistency requires \cite{Aux}
    \be
    &&[Q_{\pm}, \pet]_+|ph\hb=[Q_{\pm}, \bar{k}_{\dal}]_-|ph\hb=0
    \eel{e:618}
    or equivalently
    \be
    &&\hat{\phi}'''_{\pm}|ph\hb=\hat{\bar{d}}_{\dal}^{(\pm)}|ph\hb=0
    \eel{e:619}
    and
    \be
    &&[Q_{\pm}^2, \pet]_{-}|ph\hb=[Q_{\pm}^2, \bar{k}_{\dal}]_{-}|ph\hb=0
    \eel{e:620}
    which  implies
    \be
    &&\bar{\eta}|ph\hb=c^{\al}|ph\hb=0
    \eel{e:621}
    The physical states are then completely ghost fixed and the matter part
satisfies exactly
    the equations for the  superfields $\Phi'_{\pm}$ above. Notice also that
\r{617}, \r{619} and
    \r{621} imply \r{616}. Notice also that \r{621} makes
$Q_{\pm}^2$ vanish on physical states.

    Similarly does
    \be
    &&\bar{\pet}|ph\hb={k}_{\al}|ph\hb=0
    \eel{e:622}
    imply
    \be
    &&\eta|ph\hb=\bar{c}_{\dal}|ph\hb=0
    \eel{e:623}
    and
    \be
    &&\hat{\bar{\phi}}'''_{\pm}|ph\hb=\hat{{d}}_{\al}^{(\pm)}|ph\hb=0
    \eel{e:624}
which are exactly the equations  for the  superfields $\bar{\Phi}'_{\pm}$
above.

    \setcounter{equation}{0}
            \section{The massless superparticle models.}

 We have demonstrated that the method of ref.\cite{MM} lead in a well defined
manner
  to a gauge invariant although complex Lagrangian for the superparticle in
four
            dimensions, free or in interaction with an external super Maxwell
field. Although the
   derivation is only valid for massive superparticles the obtained
Lagrangians have
        a well defined massless limit. However, in this case we cannot
uniquely relate them to the
    massless limit of the corresponding real Lagrangians through the
generalized BRST quantization (see
    remark after \r{4019}). In fact, these two Lagrangians do not even have
the same number of degrees
    of freedom in the massless limit. (The corresponding real one has less.)
     The reason
  why our derivation is not valid in the massless limit is due to the fact
that the equations
    \r{4},\r{12} and \r{37} only follow from \r{03} and \r{34} respectively
if $m\neq0$. Below we
    consider the true massless superfield case.

    Consider first the free massless superfield. The massless limit of \r{03}
is given by
    \be
    &&D^2\Phi=0,\;\;\;{\bar{D}}^2\bar{\Phi}=0
    \eel{e:62}
    which is different from the massless limit of \r{4} and \r{12}.
Notice that $\Phi$ and $\bar{\Phi}$ are not
  chiral fields and that \r{62} cannot be derived from a chiral Lagrangian.
The corresponding gauge
    invariant complex covariant Lagrangians are \be
    &&L=L_0+vd^2,\;\;\;\bar{L}=L_0+\bar{v}\bar{d}^2
    \eel{e:63}
    where $L_0$ is given by \r{5} and where $v$ is a complex bosonic Lagrange
 multiplier. In this case
    there is no complete configuration space Lagrangians available since
    \be
    &&d^2=\pi^{\al}\pi_{\al}-2\pi(p\cdot\sigma)\bar{\theta}-{\bar{\theta}}^2
p^2,\;\;\;{\bar{d}}^2=
   \bar{\pi}_{\dal}\bar{\pi}^{\dal}-2\theta(p\cdot\sigma)\bar{\pi}-{\theta}^2
p^2
    \eel{e:64}
    which implies that the momentum $p_{\mu}$ cannot be eliminated in \r{63}.
However, $\pi^{\al}$ and
    $\bar{\pi}_{\dal}$ may be eliminated which yields
    \be
    &&L=p\cdot
    (\dot{x}-i\dot{\theta}\sigma\bar{\theta})+\frac{\dot{\theta}^2}{4v}+i
\bar{\pi}_{\dal}\dot{\bar{\theta}}^{\dal},\nn\\
    &&\bar{L}=p\cdot
    (\dot{x}+i{\theta}\sigma\dot{\bar{\theta}})+\frac{\dot{\bar{\theta}}^2}
{4\bar{v}}-
    i{\pi}^{\al}\dot{{\theta}}_{\al}
    \eel{e:65}
    $L$ yields the equations
    \be
  &&\dot{p}^{\mu}=0,\;\;\;\dot{\bar{\theta}}^{\dal}=0,
    \;\;\;\dot{x}^{\mu}=i\dot{\theta}\sigma^{\mu}{\bar{\theta}},\nn\\
    &&\dot{\bar{\pi}}_{\al}=-\dot{{\theta}}^{\al}
   (p\cdot\sigma)_{\al\dal},\;\;\;\frac{d}{dt}\left(\frac{\dot{\theta}^{\al}}
{v}\right)=0
    \eel{e:651}
   and $\bar{L}$  the complex conjugate ones. The massless condition $p^2=0$
can only be imposed as
    initial condition here as well as in \r{68} below. Notice that \eg $L$
in \r{65} is gauge invariant
    under (the nonzero transformations)
    \be
    &&\del
    x^{\mu}=i\zeta\dot{\theta}\sigma^{\mu}\batheta,\;\;\;\del\theta^{\al}=
\zeta\dot{\theta}^{\al},
    \;\;\;\del\bar{\pi}_{\dal}=-\zeta\dot{\theta}^{\al}(p\cdot\sigma)_{\al\dal}
\;\;,\;\;\delta v=\frac{d(v\zeta)}{d\tau}
    \eel{e:652}
    where $\zeta(\tau)$ is an arbitrary real infinitesimal parameter function.

    The arguments in section 4 suggests here the existence of an associated
real Lagrangian of the
    form
    \be
    &&L_{real}=L_0+vd^2+\bar{v}\bar{d}^2
    \eel{e:66}
    which after eliminating $\pi^{\al}$ and $\bar{\pi}_{\dal}$ becomes
    \be
    &&L_{real,conf}=p\cdot
    (\dot{x}-i\dot{\bar{\psi}}\ga\psi)+\frac{\dot{\theta}^2}{4v}+\frac{\dot
{\bar{\theta}}^2}{4\bar{v}}
    \eel{e:67}
    The equations of motion are
    \be
    &&\dot{p}^{\mu}=0,\;\;\;\dot{x}^{\mu}-i\dot{\bar{\psi}}\ga^{\mu}\psi=0,
    \;\;\;\dot{{\theta}}^2=\dot{\bar{\theta}}^2=0,\nn\\
    &&\frac{d}{dt}\left(\frac{\dot{\theta}_{\al}}{4v}\right)
   +i(p\cdot\sigma)_{\al\dal}\dot{\bar{\theta}}^{\dal}=0,\;\;\;\frac{d}{dt}
\left(\frac{\dot{\bar{\theta}}_{\dal}}{4v}\right)
    -i\dot{{\theta}}^{\al}(p\cdot\sigma)_{\al\dal}=0
    \eel{e:68}

    We consider now the interaction case. The massless limit of
\r{102} may be written
    \be
    &&{D}^2_{(\pm)}\Phi'_{\pm}=0,\;\;\;\bar{D}^2_{(\pm)}\bar{\Phi'}_{\pm}=0
    \eel{e:69}
    where (for simplicity we replace $g/2$ by $g$ here)
    \be
    &&D_{(\pm)\al}=D_{\al}\pm gA_{\al},\;\;\;
    \bar{D}_{(\pm)\dal}=\bar{D}_{\dal}\mp g\bar{A}_{\dal}
    \eel{e:70}
    (Notice that these equations cannot be obtained from chiral Lagrangians.)
    The corresponding gauge invariant complex covariant Lagrangians are
    \be
   &&L_{\pm}=L_0+v{d}^2_{(\pm)},\;\;\;\bar{L}_{\pm}=L_0+\bar{v}\bar{d}^2_
{(\pm)}
    \eel{e:71}
    where
    \be
    &&d_{(\pm)\,\al}=d_{\al}\pm gA_{\al},\;\;\;
    \bar{d}_{(\pm)\,\dal}=\bar{d}_{\dal}\pm g\bar{A}_{\dal}
    \eel{e:72}
    In configuration space they become
    \be
    &&L_{\pm}=p\cdot
    (\dot{x}-i\dot{\theta}
    \sigma\bar{\theta})+\frac{\dot{\theta}^2}{4v}+i\bar{\pi}_{\dal}\dot{\bar
{\theta}}^{\dal}\pm
    ig\dot{\theta}^{\al}A_{\al},\nn\\ &&\bar{L}_{\pm}=p\cdot
    (\dot{x}+i{\theta}\sigma\dot{\bar{\theta}})+\frac{\dot{\bar{\theta}}^2}
{4\bar{v}}-
    i{\pi}^{\al}\dot{{\theta}}_{\al}\mp ig\dot{\bar{\theta}}_{\dal}\bar{A}^
{\dal}
    \eel{e:73}
    where $A_{\al}$ and $\bar{A}_{\dal}$ are defined in \r{26}. $L_{\pm}$ is
gauge invariant under the
    transformations (the nonzero ones)
    \be
    &&\del
    x^{\mu}=i\zeta\dot{\theta}\sigma^{\mu}\batheta,\;\;\;\del\theta^
{\al}=\zeta\dot{\theta}^{\al},
    \;\;\;\del\bar{\pi}_{\dal}=-\zeta\dot{\theta}^{\al}(p\cdot\sigma)_
{\al\dal} \pm\zeta
    g\dot{\theta}^{\al}\partial_{\dal}A_{\al},\nn\\
    &&\del p_{\mu}=\pm i\zeta
    g\dot{\theta}^{\al}\partial_{\mu}A_{\al}\;\;,\;\;
\delta v=\frac{d(v\zeta)}
{d\tau}
  \eel{e:731} where $\zeta(\tau)$ as before is an arbitrary real
    infinitesimal parameter function.

    The corresponding real Lagrangians to \r{73} are
    \be
    &&L_{\pm(real)}=L_0+v{d}^2_{(\pm)}+\bar{v}\bar{d}^2_{(\pm)}
    \eel{e:74}
    or
    \be
    &&L_{\pm(real,conf)}=p\cdot
    (\dot{x}-i\dot{\bar{\psi}}\ga\psi)+\frac{\dot{\theta}^2}{4v}+
\frac{\dot{\bar{\theta}}^2}{4\bar{v}}\pm
    igA^{\al}\dot{\theta}_{\al}\mp ig\bar{A}_{\dal}\dot{\bar{\theta}}^{\dal}
     \eel{e:75}

    The generalized BRST quantization of section 4 applied to the Lagrangian
\r{75} leads to the
    hermitian BRST charge operators
    \be
    &&Q_{\pm}=\del_{\pm}+\del_{\pm}^{\dag},\;\;\;\del_{\pm}=
    \eta\hat{d}_{(\pm)}^2,\;\;\;\del_{\pm}^{\dag}=
\bar{\eta}\hat{\bar{d}}_{(\pm)}^2
    \eel{e:76}
    where  $\eta$ is a non-hermitian fermionic ghost. $Q_{\pm}$ are hermitian
 and  assumed   to be
    conserved. We have the properties
     \be &&\del_{\pm}^2=(\del_{\pm}^{\dag})^2=0,\;\;\;[\del_{\pm},
    \del_{\pm}^{\dag}]_+=Q_{\pm}^2=-4\eta\bar{\eta}(\hat{p}\mp2gA)_{\mu}
\sigma^{\mu}_{\al\dal}[\hat{d}_{(\pm)}^{\al},
    \hat{\bar{d}}_{(\pm)}^{\dal}]_-\neq 0
     \eel{e:77}
    where $A_{\mu}$ is given by \r{26}. $\del_{\pm}$ and  $\del_{\pm}^{\dag}$
may be viewed as  the BRST
    charges from $L_{\pm}$ and $\bar{L}_{\pm}$ in \r{71}. Consider now the
projection \r{4014}. As in the
    previous cases the genuine physical states from \r{4014} may be chosen
to satisfy
     \be
    &&\pet|ph\hb=\bar{\eta}|ph\hb=\hat{d}^2_{(\pm)}|ph\hb=0
    \eel{e:78}
    where $\pet$ and $\bar{\pet}$ are canonical conjugate momenta to the
ghosts $\eta$ and
     $\bar{\eta}$ respectively. Alternatively they may be chosen to satisfy
    \be
    &&\bar{\pet}|ph\hb=\eta|ph\hb=\hat{\bar{d}}^2_{(\pm)}|ph\hb=0
    \eel{e:79}
 Eqs. \r{78} and \r{79} correspond exactly to the equations for the massless
    superfields $\Phi_{\pm}$ and $\bar{\Phi}_{\pm}$ in
    \r{69} respectively.

\setcounter{equation}{0}
        \section{Final remarks.}

In this paper we have derived complex gauge invariant actions for
superparticles from some superfield
equations given in the literature. We have then demonstrated that the
resulting actions
correctly describe the propagators by means of a natural proper time gauge
fixing in the path
integral expressions. In fact we get the same results as in ref. \cite{FSh}
both in the free as well
as in the interaction case. The only peculiar feature of the use of complex
actions is that one has to
give up the strict reality  properties of the involved dynamical variables
both in the equations of
motion and in the gauge transformations. A similar problem was also noted for
one of the chiral
Lagrangians: The supposed reality properties of the chiral superfields in the
 real Lagrangian
describing chiral superfields in interactions with an external super Maxwell
 field are not valid in
the equations of motion.

The obtained complex gauge invariant actions were shown to be related to real
 actions with second
class constraints by means of the generalized BRST quantization proposed in
\cite{RM,RMA}. In this
way we found a new real Lagrangian (6.30) for the interaction case which is
only slightly different
from (5.6) but which like (5.6) also can be obtained by minimal coupling.
 This generalized BRST
proceedure also tell us how the BRST quantization of the gauge invariant
complex actions should be
performed: The original state space should be spanned by all dynamical
operators together with their
hermitian conjugates. To the non-hermitian nilpotent BRST charge operator
 one should add its hermitian
conjugate and perform the projection to the physical state space using both
these BRST charges. With
this procedure we have complete equivalence between the complex gauge
invariant approach and the real
approach with second class constraints. This is a little confusing in the
 massless limit of the
massive chiral cases in section 3 and 6 since the real and complex actions
 classically do not leave
the same physical degrees of freedom. Here the BRST quantization is also not
 completely equivalent to
the corresponding superfield equations due to the presence of ghost
 excitations (see remark after
(4.19)). What exactly happens in a proper BRST quantization on inner
 product spaces \cite{Propa}
remains to work out. The massless nonchiral cases considered in
section 7 do not have this problem.

            \setcounter{section}{1}
    \setcounter{equation}{0}
    \renewcommand{\thesection}{\Alph{section}}
    \newpage
    \noindent
    {\Large{\bf{Appendix}}}
     \vspace{5mm}

  In section  6 we stated that $\bar{d}_{\dal}=0$
and $\phi''_{\pm}=0$  (as well as
 $\bar{d}_{\dal}^{(\pm)}=0$ and ${\phi}'''_{\pm}=0$) are first class
constraints satisfying an
    abelian algebra. Here we give the details of this calculation.\\

    {\bf Proof of $\{\bar{d}_{\dal}, \phi''_{\pm}\}=0$.}

    	Define
     \be
       &&d_{\al}^{(\pm)}\equiv d_{\al}\pm gA_{\al}
    \eel{e:45}
    then we have
    \be
    &&\{\bar{d}_{\dal}, d_{\al}^{(\pm)}\}=2i(p\cdot\sigma)_{\al\dal}\pm ig
\bar{D}_{\dal}D_{\al}V
    \eel{e:46}
     and
    \be
    &&p_{\mu}\mp gA_{\mu}=\frac{i}{4}\bar{\sigma}_{\mu}^{\dal\al}\{\bar{d}_
{\dal}, d_{\al}^{(\pm)}\}
    \eel{e:47}
    using the conventions in \cite{WB}. $\phi''_{\pm}$ may therefore be
written as
    \be
    &&\phi''_{\pm}=m^2+\frac{1}{8}\{\bar{d}_{\dal}, d_{\al}^{(\pm)}\}\{\bar{d}
^{\dal},
    d^{\al}_{(\pm)}\}\mp \frac{g}{2}d_{\al}^{(\pm)}W^{\al}
    \eel{e:48}
    Since
    \be
    &&\{\bar{d}_{\dal}, W^{\al}\}=i\bar{D}_{\dal}W^{\al}=0\nn\\
    &&\{\bar{d}_{\dal}, \{\bar{d}_{\dot{\beta}},d_{\beta}^{(\pm)}\}\}=\pm
ig\{\bar{d}_{\dal},
    \bar{D}_{\dot{\beta}}D_{\beta}V\}=\nn\\
    &&=\mp g \bar{D}_{\dot{\al}}\bar{D}_{\dot{\beta}}D_{\beta}V=\pm
    2g\epsilon_{\dal\dot{\beta}}W_{\beta}
    \eel{e:49}
    we finally get
    \be
    &&\{\bar{d}_{\dal}, \phi''_{\pm}\}=\kvart \{\bar{d}^{\dot{\beta}}, d^
{\beta}_{(\pm)}\}(\pm
    2g)\epsilon_{\dal\dot{\beta}}W_{\beta}\mp\frac{g}{2}\{\bar{d}_{\dal},
d_{\al}^{(\pm)}\}W^{\al}=0
    \eel{e:50}\\

    {\bf Proof of $\{\bar{d}_{\dal}^{(\pm)}, \phi'''_{\pm}\}=0$.}

    \be
    &&\{\bar{d}_{\dal}^{(\pm)}, \phi'''_{\pm}\}=\{\bar{d}_{\dal},
    \phi''_{\pm}+(\phi'''_{\pm}-\phi''_{\pm})
    \}\pm\halv g \{\bar{A}_{\dal}, \phi'''_{\pm}\}=0+\{\bar{d}_{\dal},
\frac{g^2}{4}(\partial V)^2\pm\nonumber \\
&&\pm
    ig\partial V\cdot(p\mp gA)+\frac{g^2}{4}A_{\al}W^{\al}\}\pm\halv
g\{\bar{A}_{\dal},(p\mp
    gA)^2\mp\frac{g}{2}d^{(\pm)}_{\al}W^{\al}\}=\nonumber \\
    &&=i\frac{g^2}{2}\partial_{\mu}V\partial^{\mu}\bar{A}_{\dal}\mp
    g\partial\bar{A}_{\dal}\cdot(p\mp gA)+g^2\partial_{\mu}
    V\bar{D}_{\dal}A^{\mu}+i\frac{g^2}{4}\bar{D}_{\dal}A_{\al}W^{\al}-
\nonumber \\
&&-i\frac{g^2}{4}A_{\al}\bar{D}_{\dal}W^{\al}\pm
    g\partial\bar{A}_{\dal}\cdot(p\mp gA)+i\frac{g^2}{4}D_{\al}\bar{A}_
{\dal}W^{\al}=0
    \eel{e:500}
    where we have used the explicit expressions of $A^{\al},\,\bar{A}^
{\dal},\,A^{\mu}$ and $W^{\al}$ in
    terms of $V$ given in \r{26} and \r{38}.

    \end{document}